\documentclass[twocolumn]{aastex61}
\received{March 29, 2017}
\revised{April 19, 2017}
\accepted{\today}
\submitjournal{ApJ}

\shorttitle{Frequent flaring in the TRAPPIST-1 system -- unsuited for life?}
\shortauthors{Vida et al.}

\usepackage{natbib}
\usepackage{graphicx}
\usepackage{txfonts}
\usepackage{hyperref}
\usepackage{lipsum}

\newcommand{\kepler}{\textit{Kepler}}

\begin{document} 
  \title{Frequent flaring in the TRAPPIST-1 system -- unsuited for life?}

\correspondingauthor{K. Vida}
\email{vidakris@konkoly.hu}

\author{K. Vida}
\affil{Konkoly Observatory, MTA CSFK, H-1121 Budapest, Konkoly Thege M. \'ut 15-17, Hungary}
\author{Zs. K\H{o}v\'ari}
\affil{Konkoly Observatory, MTA CSFK, H-1121 Budapest, Konkoly Thege M. \'ut 15-17, Hungary}
\author{A. P\'al}
\affil{Konkoly Observatory, MTA CSFK, H-1121 Budapest, Konkoly Thege M. \'ut 15-17, Hungary}
\affil{E\"otv\"os Lor\'and  University, H-1117 P\'azm\'any P\'eter s\'et\'any 1/A, Budapest, Hungary}  
\author{K. Ol\'ah}
\affil{Konkoly Observatory, MTA CSFK, H-1121 Budapest, Konkoly Thege M. \'ut 15-17, Hungary}
\author{L. Kriskovics}
\affil{Konkoly Observatory, MTA CSFK, H-1121 Budapest, Konkoly Thege M. \'ut 15-17, Hungary}

  \begin{abstract}
   We analyze the K2 light curve of the TRAPPIST-1 system. The Fourier analysis of the data suggests $P_\mathrm{rot}=3.295\pm0.003$ days. The light curve shows several flares, of which we analyzed 42 events with integrated flare energies of $1.26\times10^{30}-1.24\times10^{33}$ ergs.  Approximately 12\% of the flares were complex, multi-peaked eruptions. The flaring and the possible rotational modulation shows no obvious correlation. The flaring activity of TRAPPIST-1 probably continuously alters the atmospheres of the orbiting exoplanets, making these less favorable for hosting life. 

\end{abstract}


\section{Introduction}
TRAPPIST-1 (2MASS J23062928-0502285) is an ultracool, M8-type dwarf, which is known to host seven terrestrial planets, with three of them having equilibrium temperatures that makes the existence of liquid water on their surface possible \citep{2016Natur.533..221G, 2017Natur.542..456G}.
The discovery obviously raised the question of the planetary habitability.
This involves several complex factors, one of them being the magnetic activity of the host star, as a large fraction of M dwarfs has been observed to be magnetically active. Numerous high energy events (e.g., flares) could erode the atmospheres of these worlds that are located very close to their host star and render them uninhabitable over time
\citep{2007AsBio...7..167K,2008SSRv..139..437Y}.
TRAPPIST-1 indeed seems to exhibit chromospheric activity as shown by its H$\alpha$ emission level \citep{2000AJ....120.1085G, 2010ApJ...710..924R}
and significant X-ray and EUV (XUV) radiation \citep{2017MNRAS.465L..74W}. 
Lyman-$\alpha$ observations suggest moderate activity, although the detected XUV radiation could be strong enough to strip the atmospheres of the inner planets in a few billions years \citep{2017A&A...599L...3B}.
In this paper we examine raw cadence K2 observations from the \kepler{} Space Telescope in order to retrieve information about its activity and impacts on habitability.

\section{Observations}
\begin{figure}
    \centering
    \includegraphics[width=0.48\textwidth]{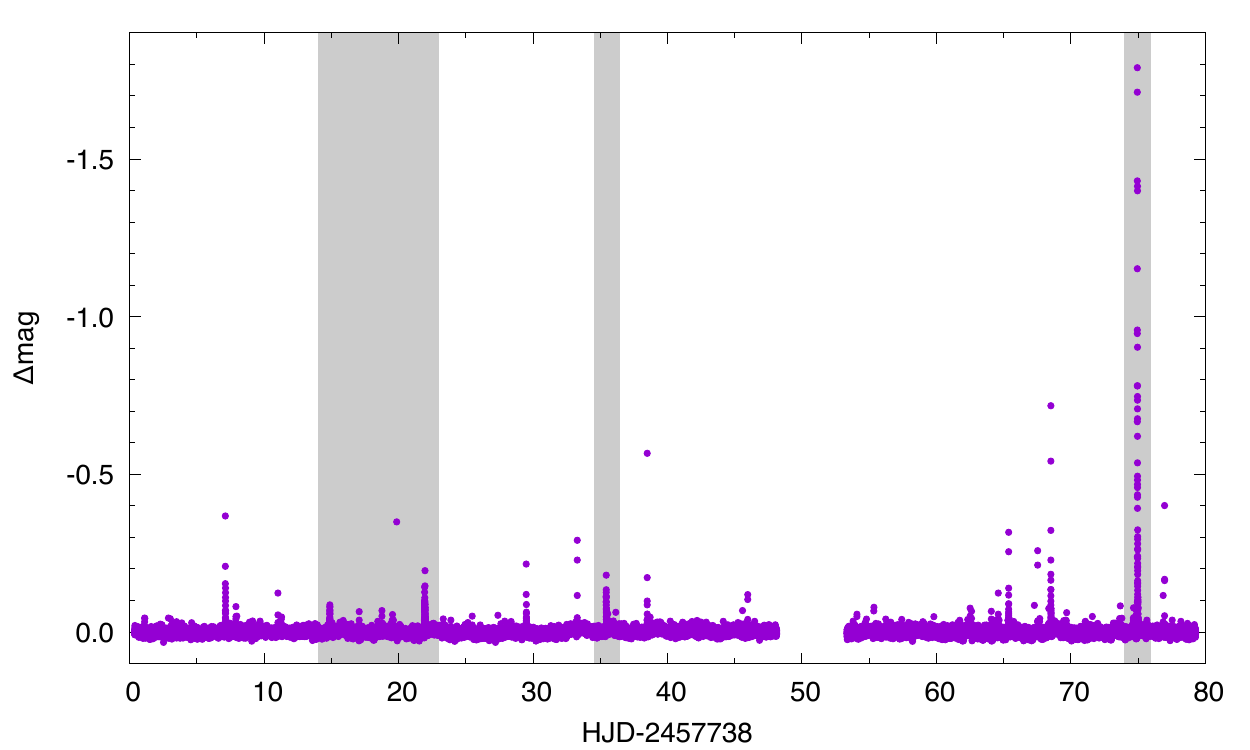}
    \caption{K2 light curve of TRAPPIST-1. Epochs where complex flare events occurred are marked with shaded areas.}
    \label{fig:lc}
\end{figure}

During this analysis, only uncalibrated (raw) K2 measurements are available
for the community. In order to analyze the K2 images and extract flux 
variations from the raw K2 data series, we converted the uncalibrated per-pixel
time series to numerous individual image files. 
These raw, per-cadence FITS files were converted to individual image stamps
by involving various tasks of the FITSH package \citep{2012MNRAS.421.1825P}. 
Although the transposed version of these data (i.e., target pixel files) has been
made available for TRAPPIST-1 almost immediately, we chose this approach
in order to have the possibility to incorporate pointing information 
retrieved from the nearby stamps. 
Handling uncalibrated data yields more prominent systematics which are inducted directly by the different sensitivity of the neighboring pixels and, hence, indirectly by the pointing jitter - which is comparable with the FWHM of the instrumental PSF in the case of K2. In the following, we detail how these systematics can be reduced while the the amplitude and shape of the flare-like events are not distorted.

In total, 107968 short cadence (SC) image stamps are available for this campaign.
For all of these frames, we computed the flux-weighted centroid of the 
target PSF and search for discontinuities and/or outliers in the $x$ and $y$ 
positions. This particular background, readout and shot noise level of the 
target star allowed us to obtain an RMS of $\sim 0.005$ pixels
in the time series of the centroid positions. We note here that this statistical
RMS agrees well with the per-image formal uncertainties of the centroid 
positions estimated directly from the aforementioned noises and the PSF shape.

By rejecting obvious cases (e.g., frames acquired during reaction wheel 
momentum resaturation phases) as well as individual lower quality images
(affected by cosmic hits, etc.), we involved 100337 SC images in the 
further analysis. We note here in this data series, that there are 216 distinct
sections where the pointing were stabilized only by the control of the two
reaction wheels. 

Analyzing and extracting flux variations related flare activity needs 
a different approach what is used in highly precise photometric 
measurements (such as searching for planetary transits). This is even crucial 
in the case of K2 where the resaturation phases occur in periods of $\sim 6$
hours which is comparable to the time scales of the flares. Hence, we 
performed a global decorrelation of the light curve by involving a
function having a part built from sine and cosine of the fractional pixel 
coordinates, namely
\begin{eqnarray}
& \sum\limits_{k=0}^N\sum\limits_{\ell=0}^N & 
\left[
A_{[x][y]k\ell}\cos(2\pi k\Delta x)\cos(2\pi \ell\Delta y)+\right.\\
& & B_{[x][y]k\ell}\cos(2\pi k\Delta x)\sin(2\pi \ell\Delta y)+\nonumber \\
& & C_{[x][y]k\ell}\sin(2\pi k\Delta x)\cos(2\pi \ell\Delta y)+\nonumber \\
& & \left.D_{[x][y]k\ell}\sin(2\pi k\Delta x)\sin(2\pi \ell\Delta y)\right].\nonumber 
\end{eqnarray}
Here $\Delta x$ and $\Delta y$ are the fractional centroid coordinates
(i.e. $0\le \Delta x,\Delta y < 1$), $[x]$ and $[y]$ refers to the 
integer part of the centroid coordinates -- and hence
$x=[x]+\Delta x$ and $y=[y]+\Delta y$. Furthermore, we extended the 
decorrelation function with additional terms proportional to the 
fitted shape parameters of the PSF (in order to decorrelate against 
changes in the effective focal distance). Of course, the unbiased
light curve RMS would even be smaller by considering a different
set of $A_{[x][y]k\ell}$, $B_{[x][y]k\ell}$,~\dots coefficients
on each section between two subsequent resaturation phases, but here we
incorporated a global fit in order to avoid any aliases from the 
similarities between flare events and trends arising from thruster usage.
In the above procedure we used an iterative sigma-clipping technique to 
perform the linear regression that yields the respective coefficients.
By analyzing the obtained light curve, we found that its sigma-clipped 
RMS is only larger by 10\% than the expected photometric uncertainty
derived from the background, readout and shot noise.

\section{Fourier analysis}
\begin{figure}
    \centering
    \includegraphics[width=0.48\textwidth]{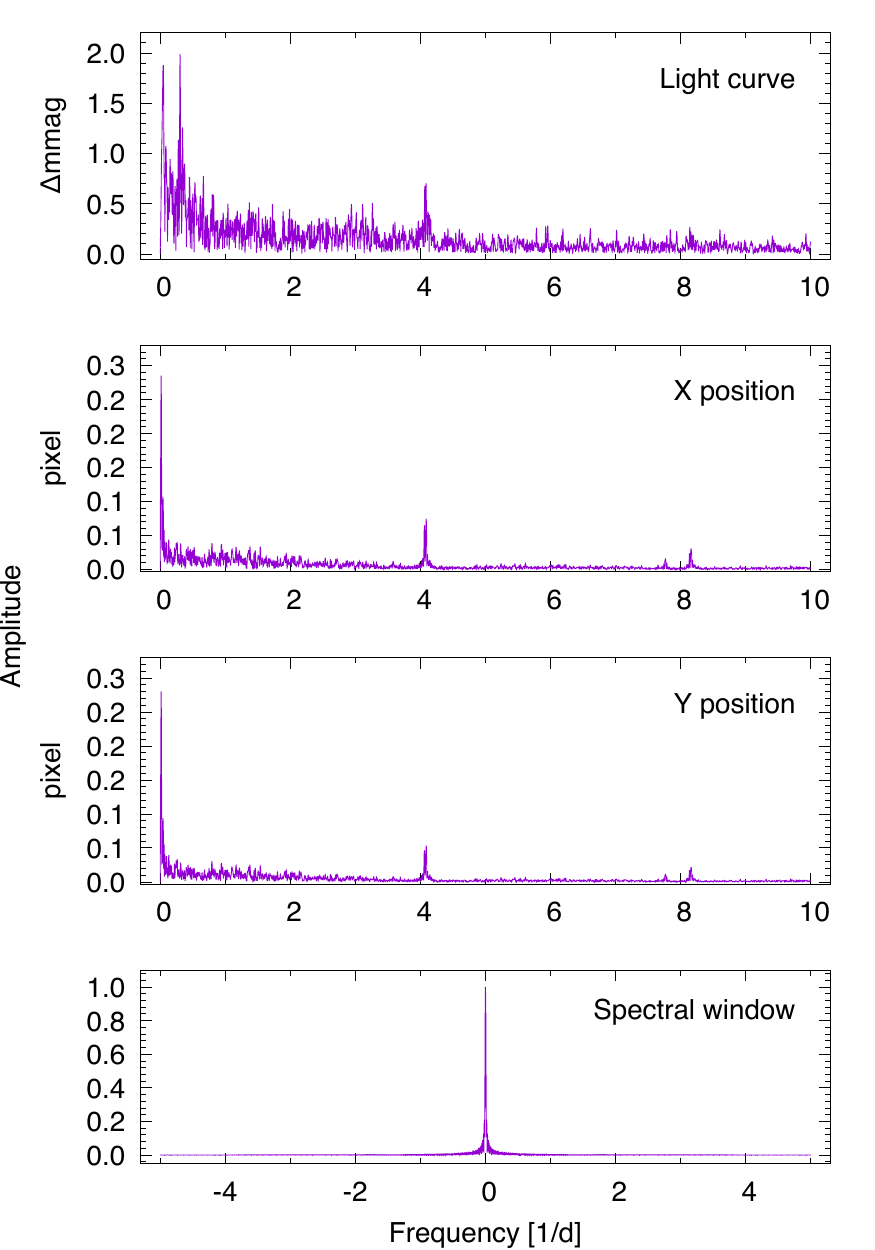}
    \caption{Fourier analysis of the cleaned and binned data. Top plot shows the Fourier spectrum of the light curve, the middle plots show the same analysis performed on the $x$ and $y$ coordinates of TRAPPIST-1 on the CCD to reveal instrumental artifacts (see text for details). The spectral window is show in the bottom plot.}
    \label{fig:fourier}
\end{figure}

Fourier analysis of the light curve was performed using MUFRAN\footnote{\url{http://www.konkoly.hu/staff/kollath/mufran.html}} \citep{mufran} 
-- a code that can do discrete Fourier transformation of data and pre-whiten light curves with the detected frequencies. 
For the analysis we used the detrended dataset manually cleaned from flares and binned to 0.05 days.
The resulting Fourier spectra and spectral window are plotted in Fig. \ref{fig:fourier}.

One important feature of the K2 data is the presence of instrumental trends. 
Without stabilization around the third axis, the instrument slowly rolls about its optical axis which causes a rotation of the field of view. This is corrected by the on-board thrusters in $\approx$6-hour periods. These corrections cause artifacts in the Fourier spectrum at 5.87 and 5.91 hours ($f_\mathrm{corr}=4.085180$ and $4.068396$\,days$^{-1}$). Our values somewhat differ from artifacts in earlier data (e.g. \citealt{2015ApJ...804L..45P}), as the corrections have been refined over time.
The artifacts can be identified by performing the Fourier analysis on the $x$ and $y$ coordinates of the target (also shown in Fig. \ref{fig:fourier}).

The most significant peak in the light curve is at $f= 0.303469$\,days$^{-1}$, $P_1=3.295\pm0.003$ days, as also found by \cite{2017arXiv170304166L}. This probably corresponds to the rotation period of the spotted star. 
Interestingly, \cite{2016Natur.533..221G} identified 
$P_\mathrm{rot}=1.40\pm0.05 d$ from ground-based TRAPPIST photometry. This latter value is consistent with the  measured 
$v\sin i=6\pm2$~km\,s$^{-1}$ \citep{2010ApJ...710..924R}, that yields 
$P_\mathrm{rot}=0.9866 d$ (0.74--1.48 days with the given error)
using $R=0.117R_\odot$ \citep{2016Natur.533..221G} and assuming $i=90^\circ$. However, our Fourier spectrum of the K2 light curve does not show any significant feature indicating a similar rotation period.

After pre-whitening the light curve with this signal, a weaker peak remains with $f= 0.342994$\,days$^{-1}$, $P_2=2.915$ days. Such signals nearby the rotational frequency are often associated with differential rotation, however, this would yield to a strong surface shear $({P_1-P_2})/{\overline{P}}$ of $\approx$0.12, which is incompatible with such a rapidly rotating and actually fully convective M dwarf.

Two longer signals are also present, corresponding to $22.3$ and $37.5$ days, but from the $x$ and $y$ coordinates of the star we get similar signals of 25--26 days as well, therefore the reality of these periods are questionable.

\section{Flares in the light curve}
\begin{figure}
    \centering
    \includegraphics[width=0.41\textwidth]{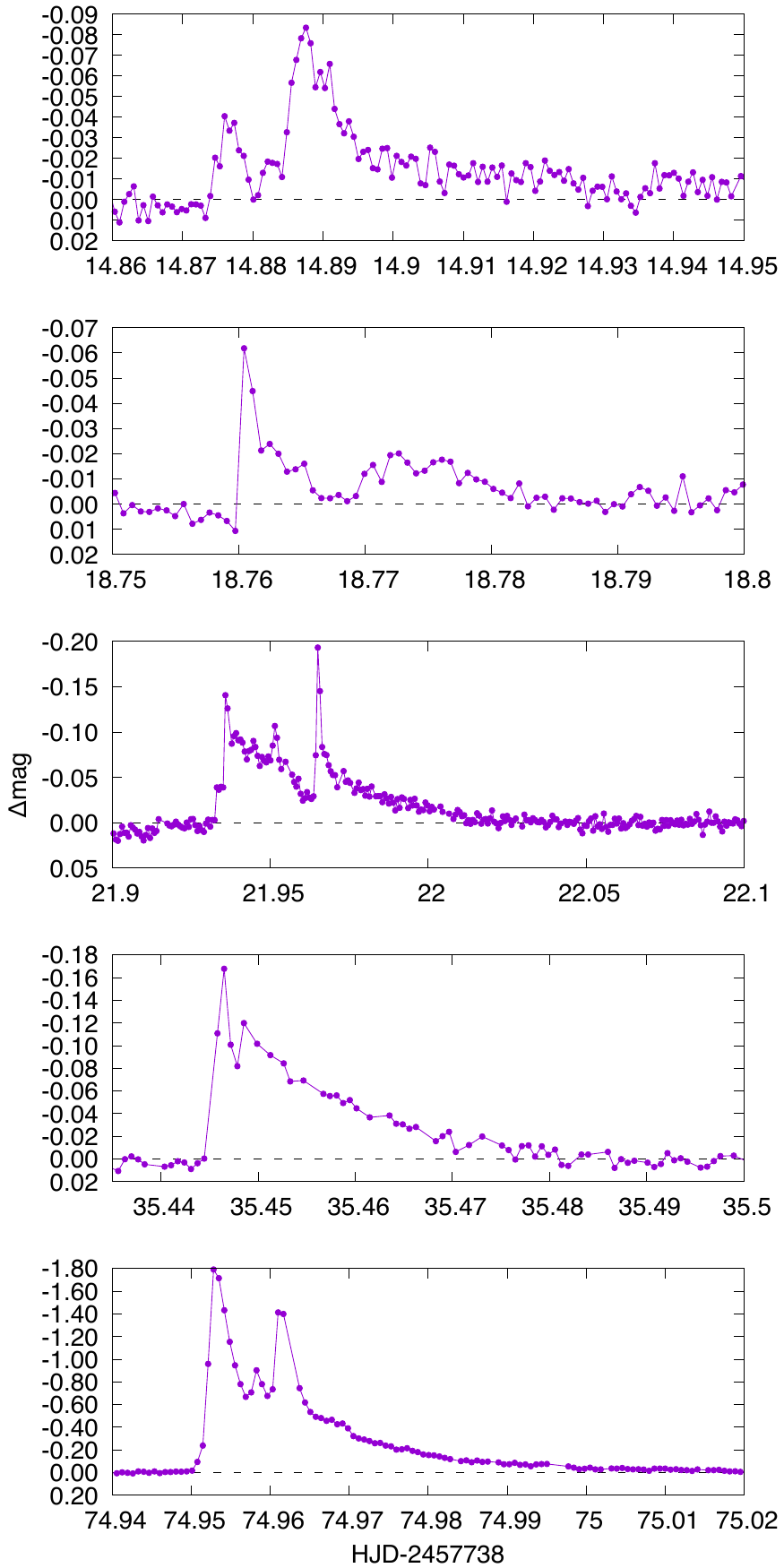}
    \caption{Light curves of the five complex flare events.}
    \label{fig:lcflare}
\end{figure}
The K2 light curve indicates strong flaring activity on TRAPPIST-1. We identified 42 flare events in the data by visual inspection of the detrended light curve. Of these events, 5 (12\%) were complex, multi-peaked eruptions, plotted in Fig. \ref{fig:lcflare}. This ratio is quite similar to that found in the much larger sample of the M4-type GJ~1243 \citep{2016ApJ...829..129S}.

\subsection{Flare energies}
\begin{figure}
    \centering
    \includegraphics[width=0.48\textwidth]{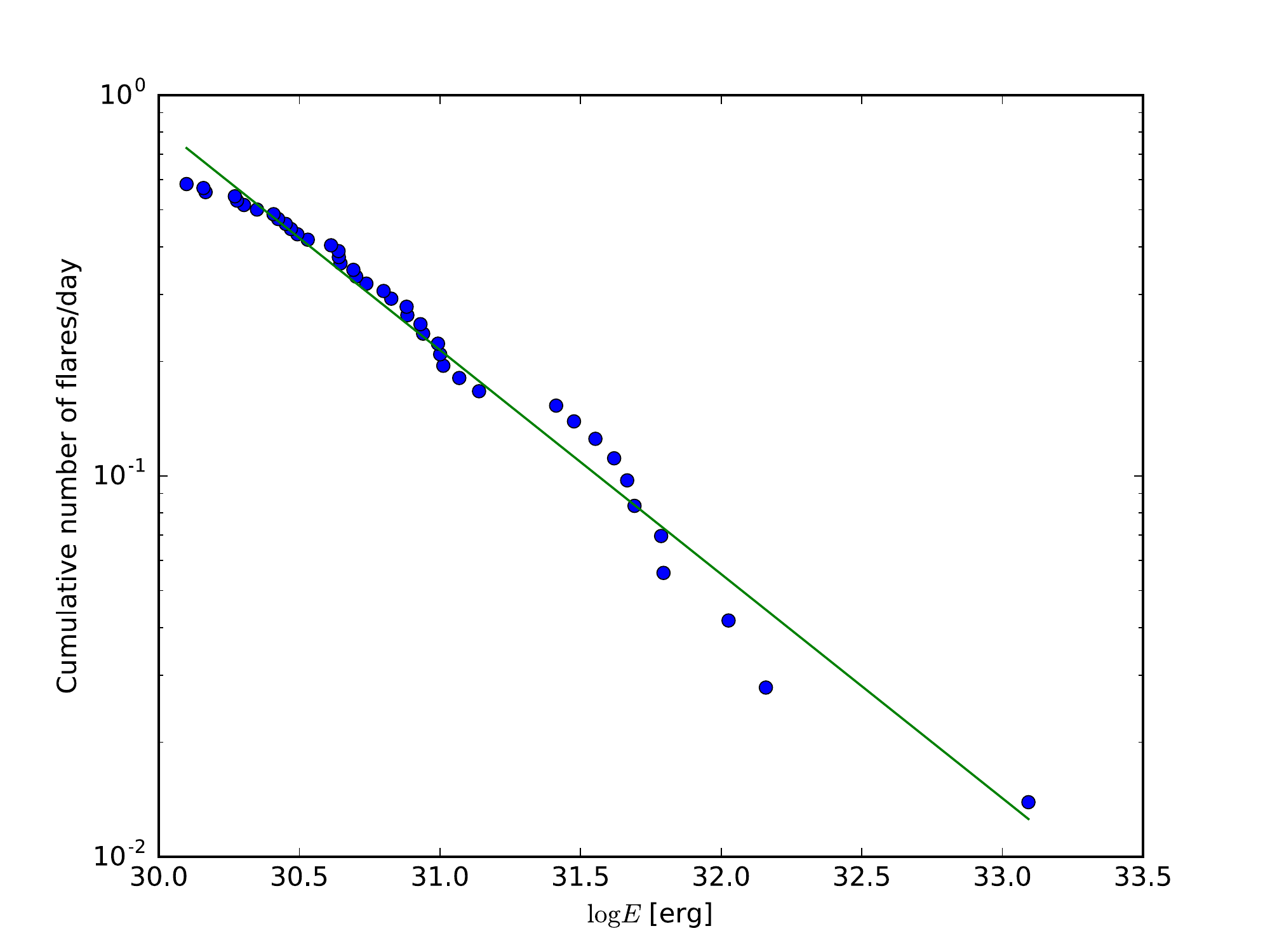}
    \caption{Cumulative flare frequency distribution fitted by a linear function}
    \label{fig:flarestat}
\end{figure}

The energies of these events were estimated following the method of \cite{2007AN....328..904K}. This method is based on integrating the flare intensity over its duration, which yields the relative flare energy (also often mentioned as equivalent duration in the literature):
$$ \varepsilon_f=\int_{t_1}^{t_2} \left( \frac{I_{0+f}(t)}{I_0}-1  \right)dt,$$
where the intensity is calculated from the magnitude:
$$\frac{I_{0+f}(t)}{I_0} = 10^{\frac{\Delta m_{Kp}}{2.5}}.$$ 
Here, $I_{0+f}$ and $I_0$
are the intensity values of the flaring and the quiescent stellar surfaces, respectively.
From the relative flare energy, $\varepsilon_f$, the integrated energy of the flare event can be calculated by multiplying it by the quiescent stellar flux:
$$ E_f=\varepsilon_f F_\star.$$
The quiescent flux can be estimated if we assume black body radiation from the effective temperature $T_\mathrm{eff}=2550$ K and radius $R=0.117R_\odot$ \citep{2016Natur.533..221G}:
$$F_\star = \int_{\lambda_1}^{\lambda_2} 4\pi R^2 \mathcal{F}(\lambda)S_{Kp}(\lambda)d\lambda .$$
Here, $\mathcal{F}(\lambda)$ is the power function and $S_{Kp}(\lambda)$ is the \kepler{} response function.
The derived energies and their occurrence rates are plotted in Fig. \ref{fig:flarestat}. 
The detection of the smallest flare events is limited by the light curve scatter (typically $\approx 0.01$ magnitude) and, mainly, the sampling of the data ($\approx 59$ seconds), thus, the weakest events appear only as a single measurement point (these were not analyzed as they could be of instrumental origin). The weakest analyzed event in our study has $E_f=1.26\times10^{30}$\,erg, while the strongest eruption, having a $Kp=1.78$\,mag peak, produced $E_f=1.24\times10^{33}$ erg energy.

If $dN$ is the number of flares in the energy range $E+dE$ then the following power law can be written
$$  dN(E) \propto E^{-\alpha}dE, $$
\citep[see e.g.][and the references therein]{2014ApJ...797..121H}. By integrating, the cumulative flare frequency distribution can be expressed in logarithmic form as
$$ \log\nu = a+\beta\log E$$
where $\nu$ is the cumulative frequency of flares with a given energy larger than $E$, while $\beta=1-\alpha$ \citep{2017arXiv170308745G}. This relationship, seen in Fig. \ref{fig:flarestat}, can be fitted by a linear function, giving the slope $1-\alpha$, where $\alpha$ is often used to characterize how the flare energy is dissipated. The best fit yields $\alpha=1.59$, suggesting that TRAPPIST-1 flare energies are mostly nonthermal \citep[cf.][]{2016ApJ...832...27A}, similarly to other very active M dwarfs in the \citet[][]{2014ApJ...797..121H} sample.
The flare energies detected on TRAPPIST-1 are somewhat higher to those found on the other nearby planet-hosting M5.5-class Proxima Centauri, where \cite{2016ApJ...829L..31D} reported eruptions with $10^{29}-10^{31.5}$\,erg energies based on MOST observations, with a similar flaring activity characterized by $\alpha=1.68$.

\subsection{Possible connection between flares and spotted regions}
\begin{figure}
    \centering
    \includegraphics[width=0.50\textwidth]{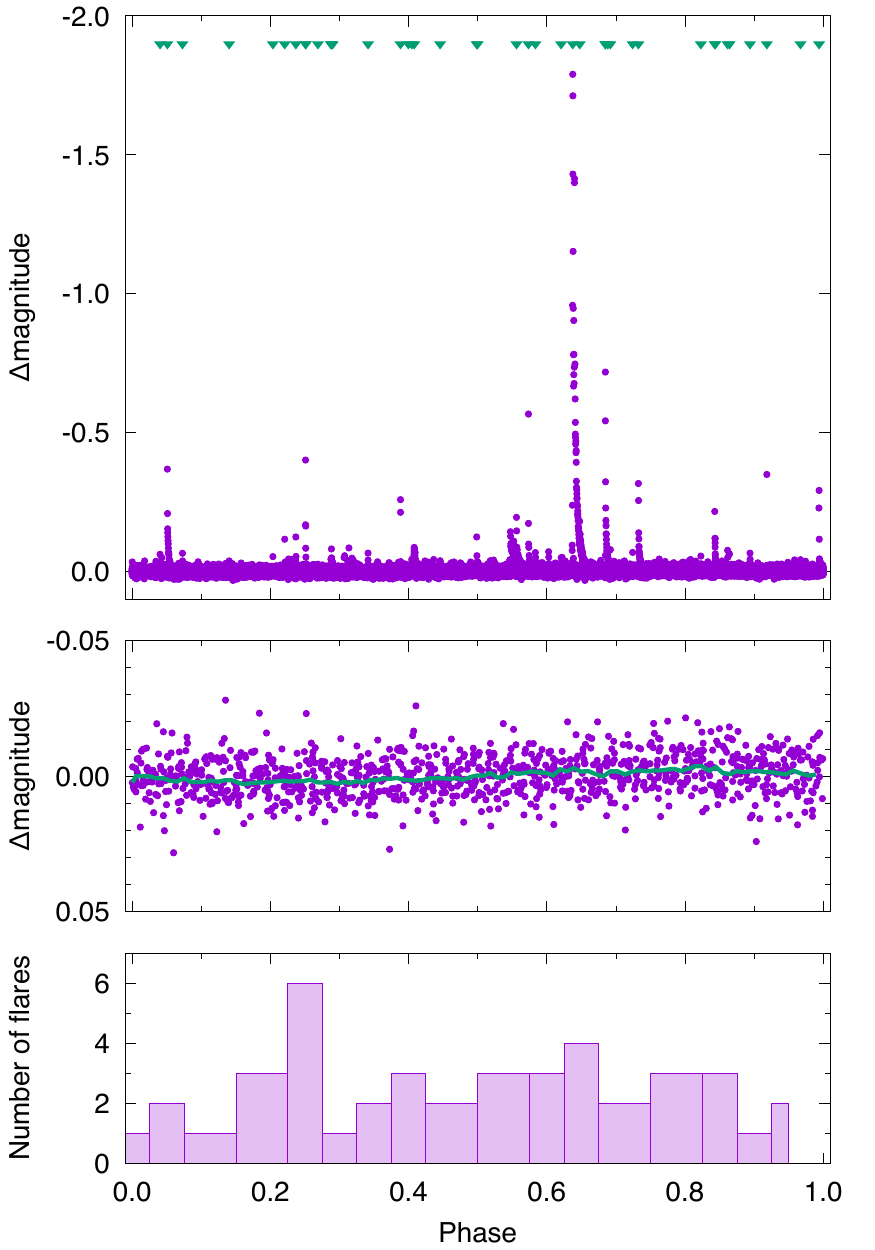}
    \caption{\textit{Top:} phased light curve of TRAPPIST-1. Triangles mark the phases of analyzed flare events. \textit{Middle:} phases light curve of TRAPPIST-1 showing the possible rotational modulation of 3.295 days (every 100 points plotted). Green line shows a median of the data. \textit{Bottom:} histogram showing the flare occurrence rate in different phases.}
    \label{fig:phase}
\end{figure}

If we accept the dominant signal in the Fourier spectrum ($P_1=3.295\pm0.003$ days) as a modulation caused by the rotation of the spotted surface, we can evaluate if there is any connection between cool surface spots and flaring activity that could indicate a connection of photospheric and chromospheric activity, as seen on the Sun and also found on other stars,  e.g. on the BY Dra-type EY Dra \citep{2010AN....331..772K}
or V374 Peg \citep{2016A&A...590A..11V} 
and other BY Dra-type stars \citep[e.g.]{2005AJ....130.1231P};
on the K-dwarf 
PW And \citep{2003A&A...411..489L};
RS CVn-type binaries, e.g. 
II Peg, and $\lambda$ And \citep{2008A&A...479..557F};
and on the W UMa-type VW Cep \citep{1996A&A...313..532F}.  

In Fig. \ref{fig:phase} we plotted the light curve phased with the following ephemeris:
$$\mathrm{HJD} =2457738.362703 + 3.295\times E.$$
With this phasing we find that flares can be found at every phase, although they are somewhat more frequent at the light curve minimum (around phase 0.25), i.e, where the surface spottedness is higher. Interestingly, the strongest flares seem to appear around the light curve maximum (phases 0.55--0.75). This behaviour is very similar to the somewhat hotter, but still fully convective M4-type V374\,Peg \citep{2016A&A...590A..11V}.

\subsection{The complex flare event at HJD 2457812}
\begin{figure}
    \centering
    \includegraphics[width=0.50\textwidth]{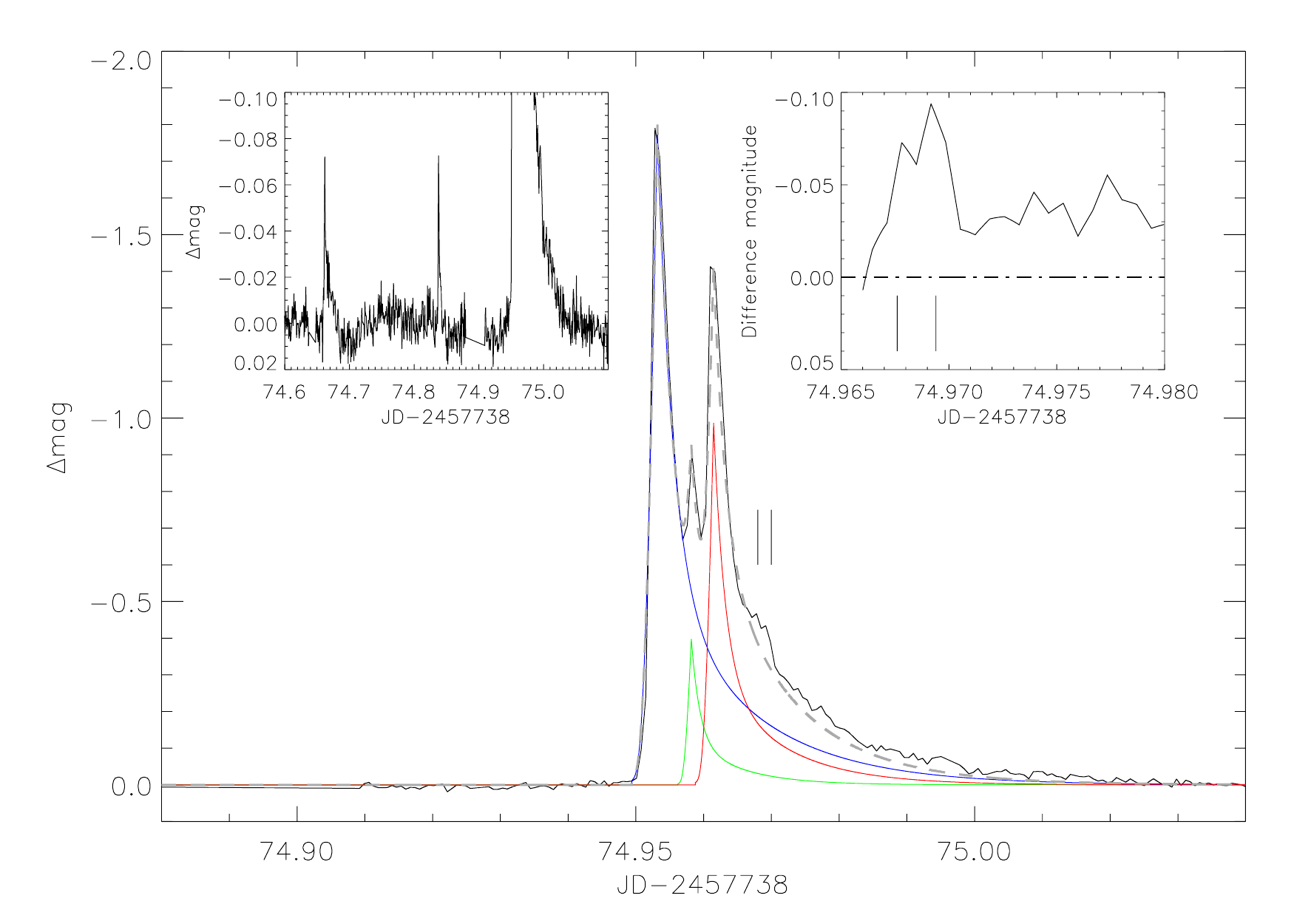}
    \caption{The complex flare event on HJD 2457812. The inset on the left shows the two earlier eruption that could be connected with the large flare, while the one on the right shows two further possible eruptions after subtracting an analytical fit. Each model is plotted with different color. The sum is plotted with a dashed grey line. See text for further details.}
    \label{fig:largeflare}
\end{figure}
The light curve shows several complex flare events (shown in Fig. \ref{fig:lcflare}), of these the largest occurred at HJD 2457812 (on day 74 in the light curve). This event consisted of three large eruptions, and at least two smaller ones. 
We fitted the event with the sum of three separate analytical flare models following the method and parametrization described in \cite{2014ApJ...797..122D}, shown in Fig. \ref{fig:largeflare}. After the subtraction of the fit from the light curve, two smaller eruptions are suspected on the decay phase of the complex event (see the inset in Fig. \ref{fig:largeflare}).

It is possible, that this complex flare was triggered by an earlier event. Such events, called sympathetic eruptions, are often seen on the Sun, but they were observed also on other stars (e.g. \citealt{2016A&A...590A..11V} and references therein). 
Since the median waiting time between consecutive flares on TRAPPIST-1 is 28.1 hours, and the two smaller flares occurred approximately 7.7 and 3.4 hours earlier before the main eruption, they are likely to be connected with each other.
One interesting similar example that is known in details is the 2010 August event on the Sun, which was modelled by \cite{2011ApJ...739L..63T}. 
According to the model, these sympathetic flares are triggered by nearby eruptions in an appropriate magnetic field structure of the corona.

\section{Discussion}
   
\cite{2016ApJ...830...77V} examined the possible influence of stellar flares on the orbiting exoplanets and their atmosphere. The basis of their analysis was the great flare on AD Leo in 1985 observed 
in high detail with both UV spectroscopy and high temporal resolution multi-passband photometry \citep{1991ApJ...378..725H}. 
That complex flare event on AD Leo, which has by chance very similar light curve shape to the largest flare seen in the K2 light curve of TRAPPIST-1, has an amplitude of $V\approx0.5$ magnitude. \cite{2016ApJ...830...77V} concluded, that atmosphere of the two super-Earth-like hypothetical planets, that would orbit AD Leo, would be altered irreversibly and significantly after such eruption. Their model suggests that the post-flare steady state would only return on the scale of $\approx30000$ years, thus, the planetary atmosphere would be constantly altered by eruptions due to the high flaring rate. Although the flaring frequency on TRAPPIST-1 is somewhat lower than on AD Leo, the flaring rate derived from the K2 data suggests that the planetary atmospheres in the TRAPPIST-1 systems would not have a steady state, which is disadvantageous for hosting life.

These eruptions pose a threat to habitability, since during a flare the UV radiation level is also increased, that can both erode planetary atmospheres on the long term, and directly harm life on the surface.
\cite{2010AsBio..10..751S} suggested that stellar flares do not necessarily affect directly the habitability, as much of the UV radiation can be absorbed by photochemical reactions in the stratosphere (supposing an Earth-like atmosphere), which would prevent it from reaching the planetary surface. 
However, the authors also suggest that due to the repeated flare events the atmosphere of the planet can be continuously disturbed -- these long-term effects still need to be understood.
Such UV absorbers, like ozone in the planetary atmosphere, can reduce the negative effects of the flares  -- the detection of such atmospheric features would suggest that the planets are more likely to have habitable environments \citep{2015ApJ...809...57R, 2017arXiv170206936O}.

We can also do a rudimentary estimation how such eruption changes temporarily the limits of the habitable zone based on the model of \cite{2013ApJ...765..131K}. In quiescent state of TRAPPIST-1, the conservative habitable zone for a 1 Earth mass planet spreads between 0.024--0.049 AU ($T_\mathrm{eff}=2550$K, $L/L_\odot=0.000525$). During the flare, supposing 1--1.5 magnitude brightening of the star, resulting in an increased luminosity of $L/L_\odot=$0.0013--0.0021. These values yield to the habitable zone limits of 0.038--0.077  and 0.048--0.097 AU respectively, a very significant change. 
Of course, the short time scale of a flare is too brief to significantly change surface temperatures of the planets and actually shift the habitable zone boundaries, 
 although the possible cumulative effect of flaring may still change the habitable zone boundaries compared to the conservative models which take into account only the quiescent stellar flux.
Our simple estimation illustrates well the magnitude of energy changes in the planetary system caused by the increased stellar energy output during a strong flare.
This very crude calculation also indicates, that habitability around TRAPPIST-1 and similar late-type dwarfs could be questionable. However, further spectroscopic observations of the TRAPPIST-1 system could help us to understand how the magnetic activity of the host star influences planetary atmospheres and possibly interact with their magnetic field.

It is possible however, that a sufficiently strong magnetic field of the orbiting exoplanet could shield the atmosphere from the harmful effects of such eruptions. \cite{2016ApJ...826..195K} found that for typical coronal mass ejection (CME) masses and speeds measured on M dwarfs, the orbiting rocky planets would need magnetic fields between tens to hundreds of Gauss, while hot Jupiters would only require magnitudes between a few and 30\,G. The authors concluded that rocky exoplanets possibly could not generate sufficient magnetic field to shield their atmosphere from these eruptions (e.g., Earth exhibits a magnetic field of $\approx 0.5\,\mathrm{G}$). 
According to \cite{2013A&A...557A..67V} planetary magnetic fields should be stronger. An Earth-like planet in the habitable zone around an M-dwarf should have magnetic fields of the order of $\approx10-10^3$ Gauss in order to possess a magnetosphere comparable to that of the present-day Earth.

Moreover, typical solar flares release energies of the order of $10^{27}-10^{32}$\,erg. One of the most energetic event, known as the 'Carrington Event' in 1859 released about $10^{33}$\,erg energy and resulted in one of the strongest
geomagnetic storms reaching the surface of the Earth. Such powerful flares (and possibly related CMEs) on TRAPPIST-1 occur more often, and hit the planetary surfaces from much shorter distances of $\approx0.011-0.063$\,AU \citep{ 2017Natur.542..456G}. This implies that magnetic storms in the TRAPPIST-1 exoplanetary system can be $10^2-10^4$ times stronger than the most powerful geomagnetic storms on Earth, which also would question the existence of a complex, highly organized life in this system.

The flaring activity of the TRAPPIST-1 system does not necessarily rule out the existence of life, but it raises doubts of hosting life as we know it on Earth. The conditions in the TRAPPIST-1 system are probably rather hostile for an Earth-like biosphere, but even on Earth, there are organisms that can tolerate extreme conditions. It is possible that biota of the TRAPPIST-1 system can survive the activity of the host star by living underground or underwater, or e.g. by using photoprotective biofluorescence as theoretized by \cite{2016arXiv160806930O}. However, the constantly changing planetary atmospheres will make the detection of biomarkers much more difficult.

These questions on habitability are important, since the estimated age of TRAPPIST-1, which is at least 500 Myr \citep{2015ApJ...810..158F} (\citealt{2017arXiv170304166L} estimated 3--8\,Gyr) could make possible the formation of life in ideal conditions, as the earliest life on Earth dates back to approximately 4\,Gyrs \citep{firstlife}, although complex life took much longer time to form (see \citealt{multicell} and references therein).

\section{Summary}

   \begin{itemize}
      \item The Fourier spectrum of the light curve indicates a possible rotation period of $P_\mathrm{rot}=3.295\pm0.003$ days;
      \item The light curve shows several flares with integrated flare energies of $1.26\times10^{30}-1.24\times10^{33}$ ergs, of these $\approx12\%$ are complex, multi-peaked; 
      \item We did not find an obvious correlation between the spottedness and flaring activity;
      \item The frequent strong flares of TRAPPIST-1 are probably disadvantageous for hosting life on the orbiting exoplanets, as the atmospheres of the exoplanets are constantly altered and cannot return to a steady state, however, this magnetic activity does not necessarily rule out the possibility of life.
   \end{itemize}

\begin{acknowledgements}
We thank an anonymous referee for helpful comments and suggestions.
The authors thank the useful discussion with L. S\'agi.
This work has used K2 data from the proposal number GO12046. Funding for the Kepler and K2 missions is provided by the NASA Science Mission directorate. 
The authors acknowledge the Hungarian National Research, Development and Innovation Office
grants OTKA K-109276, OTKA K-113117, and supports through
the Lendület-2012 Program (LP2012-31) of the Hungarian Academy of Sciences,
and the ESA PECS Contract No. 4000110889/14/NL/NDe. 
KV is supported by the Bolyai J\'anos Research Scholarship of the Hungarian Academy of Sciences.

\end{acknowledgements}

\bibliographystyle{aa} 

\end{document}